\documentclass[conference,a4paper]{APSIPA2025}
\usepackage{amsmath}
\usepackage{graphicx}
\usepackage{multirow}
\usepackage{amssymb,epsfig}
\usepackage{threeparttable}
\usepackage[backend=biber,style=ieee,]{biblatex}
\usepackage{geometry}
\usepackage{fancyhdr}
\fancypagestyle{firststyle}{
  \fancyhf{}
  \fancyhead[C]{2025 Asia Pacific Signal and Information Processing Association Annual Summit and Conference (APSIPA ASC)}
}

\begin{document}

\title{Non-Intrusive Intelligibility Prediction for Hearing Aids: Recent Advances, Trends, and Challenges}

\author{
\authorblockN{
Ryandhimas E. Zezario
}

\authorblockA{
Research Center for Information Technology Innovation, Academia Sinica, Taipei, Taiwan \\
E-mail: ryandhimas@citi.sinica.edu.tw}

}

\maketitle
\thispagestyle{firststyle}
\pagestyle{fancy}

\begin{abstract}
This paper provides an overview of recent progress in non-intrusive speech intelligibility prediction for hearing aids (HA). We summarize developments in robust acoustic feature extraction, hearing loss modeling, and the use of emerging architectures for long-sequence processing. Listener-specific adaptation strategies and domain generalization approaches that aim to improve robustness in unseen acoustic environments are also discussed. Remaining challenges—such as the need for large-scale, diverse datasets and reliable cross-profile generalization—are acknowledged. Our goal is to offer a perspective on current trends, ongoing challenges, and possible future directions toward practical and reliable HA-oriented intelligibility prediction systems.
\end{abstract}

\section{Introduction}
Speech intelligibility has long been a key metric for evaluating hearing aid (HA) performance, aiming to estimate how well audio signals can be understood by listeners. While human-based evaluations are considered the gold standard due to their reliability, they require a sufficient number of listeners to obtain unbiased evaluation scores. Conventional speech intelligibility methods often rely on signal processing and psychoacoustic models to approximate human auditory perception. Prominent examples include the Speech Intelligibility Index (SII) \cite{ref_36}, Extended SII (ESII) \cite{ref_37}, Speech Transmission Index (STI) \cite{ref_38}, Short-Time Objective Intelligibility (STOI) \cite{ref_39}, Modified Binaural STOI (MBSTOI) \cite{ANDERSEN20181}, and the Hearing Aid Speech Perception Index (HASPI) \cite{katehaspi}. Although these methods have demonstrated notable performance, they depend on the availability of clean reference signals for more accurate estimation, limiting their applicability in real-world scenarios where such references are often unavailable.

With the advancement of deep learning, there has been growing interest in applying neural network models for non-intrusive speech intelligibility assessment \cite{andersen2018nonintrusive, ref_52, mosa-net, zezario2022mti, TMINT-QI}. Depending on the type of ground-truth labels employed, deep learning-based non-intrusive intelligibility prediction methods can be categorized into objective-based and subjective-based approaches. In objective-based methods, models are trained to predict scores from objective intelligibility metrics such as STOI or HASPI. In contrast, subjective-based methods use human listening test results as target labels. Early efforts in this field primarily focus on predicting intelligibility for normal-hearing listeners. For example, STOI-Net \cite{ref_52} uses a convolutional neural network (CNN) and a bidirectional long short-term memory (BLSTM) architecture to estimate STOI scores from speech features in a non-intrusive manner. In the subjective-based domain, one of the earliest models \cite{andersen2018nonintrusive} employs CNN to predict intelligibility scores collected from human listeners. More recently, multi-target models such as MTI-Net \cite{zezario2022mti} combine CNN-BLSTM architectures with diverse acoustic features—including spectral features, raw waveforms, and self-supervised representations—to jointly predict multiple intelligibility metrics, including human ratings, STOI, and word error rate (WER), using a multi-task learning framework.

\graphicspath{ {./images/} }
\begin{figure}[t]
\centering
\includegraphics[width=8.3cm]{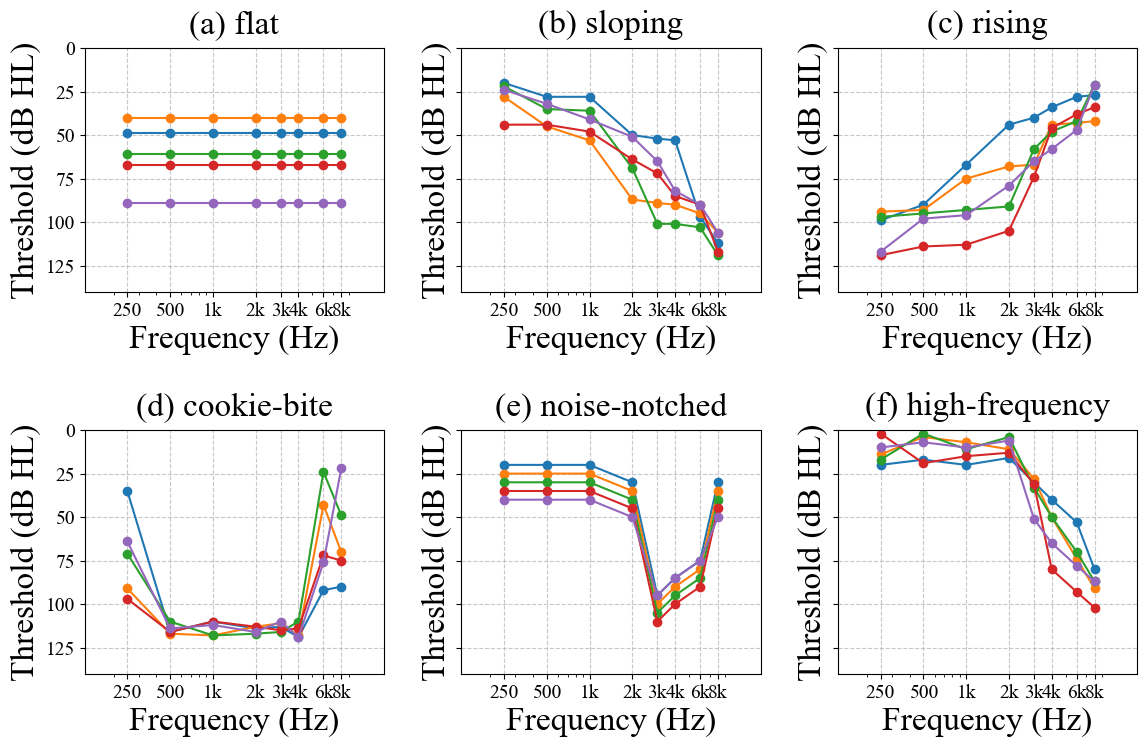} 
\caption{Examples of Hearing Loss Audiogram Profiles.} 
\label{fig:concat}
\end{figure}

Building on the promising results achieved in normal-hearing intelligibility prediction, recent studies \cite{chiang2021hasa, barker2022is, barker2024, tu22_interspeech, zezario2022mbi, close22_interspeech, chiang2023multiobjective, cuervo2024speech, zezario24_interspeech, MAWALIM2023109663} have begun to explore applications in HA scenarios, where accurate intelligibility assessment becomes even more critical. Unlike normal-hearing conditions, intelligibility for hearing-impaired (HI) listeners is influenced by a wide range of factors, including the severity of hearing loss, the characteristics of the HA processing pipeline (e.g., noise reduction, beamforming, dynamic range compression), and environmental conditions. This added complexity makes intelligibility assessment more challenging but also more clinically relevant. As such, there is a growing need for models that can generalize to diverse listening profiles and HA systems. In this survey, we provide a comprehensive overview of recent advances, highlight emerging trends in model design and evaluation, and discuss key challenges in developing reliable non-intrusive intelligibility prediction models for HA.

\section{Recent Advances and Trends}
\begin{table}[t]
\centering
\caption{Audiogram Thresholds (in decibels hearing level(dB HL).)}
\begin{tabular}{|c|c|c|c|c|c|c|}
\hline
\textbf{Frequency (Hz)} & 250 & 500 & 1000 & 2000 & 4000 & 8000 \\
\hline
\textbf{Right Ear} (dB HL) & 10 & 15 & 20 & 25 & 30 & 35 \\
\hline
\textbf{Left Ear} (dB HL)  & 15 & 20 & 25 & 30 & 35 & 40 \\
\hline
\end{tabular}
\label{tab:audiogram}
\end{table}

\begin{table*}[h]
\centering
{\normalsize
\caption{Comparison of non-intrusive intelligibility prediction methods with audiogram and acoustic feature integration.}
\resizebox{\textwidth}{!}
{%
\begin{tabular}{p{4.2cm}p{4.2cm}p{4.2cm}p{4.2cm}p{4.2cm}}
\hline
\textbf{Model / Paper (Year)} & \textbf{Acoustic Input} & \textbf{Audiogram / HL Use} & \textbf{Architecture} & \textbf{Loss / Objective} \\
\hline
Chiang et al. (2021)~\cite{chiang2021hasa} & Spectral features + audiogram (concatenated) & Direct input embedding & BLSTM + Attention & Utterance-level MSE + Frame-level MSE \\
Close et al. (2022)~\cite{close22_interspeech} &Spectral features & MSBG simulation & CNN & Utterance-level MSE \\
Tu et al. (2022)~\cite{tu22_interspeech} & ASR-based intermediate features & MSBG simulation & Transformer & CTC loss \\
Zezario et al. (2022)~\cite{zezario2022mbi} & WavLM or HuBERT & MSBG simulation & CNN + BLSTM + Attention & Utterance-level MSE + Frame-level MSE \\
Robach et al. (2022)~\cite{robach22_interspeech} & MFCC & None & TDNN & Lattice-free maximum mutual information
(LF-MMI) + Cross Entropy \\
Titalim et al. (2022)~\cite{9980000} & WavLM+ MFCC + eGeMAPS & EarModel simulation & CNN + WaveNet + Meta regressor + Stack regressor & Utterance-level MSE \\
Chiang et al. (2023)~\cite{chiang2023multiobjective} & WavLM + Audiogram (concatenated) & Direct input embedding & BLSTM + Attention & Utterance-level MSE + Frame-level MSE \\
Mawalim et al. (2023)~\cite{MAWALIM2023109663} & WavLM+ MFCC + eGeMAPS & EarModel simulation & CNN & Utterance-level MSE \\
Mawalim et al. (2023)~\cite{10289742} & WavLM+ MFCC + eGeMAPS & EarModel simulation & CNN + Stack regressor & Utterance-level MSE \\
Close et al. (2023)~\cite{close2023nonintrusiveintelligibilitypredictor} & SSL & None & BLSTM + Attention & Utterance-level MSE \\
Mogridge et al. (2024)~\cite{mogridge2024nonintrusive} & Whisper + Whisper exemplar & None & BLSTM + Attention & Utterance-level MSE \\
Cuervo et al. (2024)~\cite{cuervo2024speech} & Whisper + Audiogram (concatenated) & Direct input embedding & Transformer & Huber loss \\
Zezario et al. (2024)~\cite{zezario24_interspeech} & Spectral + Waveform + Whisper & MSBG simulation & CNN + BLSTM + Attention & Utterance-level MSE + Frame-level MSE + Cross Entropy \\
Zhou et al. (2025)~\cite{10870355} & WavLM & MSBG simulation & LSTM + LightGBM & Utterance-level MSE \\
Yamamoto et al. (2025)~\cite{yamamoto2025nonintrusivebinauralspeechintelligibility} & Whisper + audiogram & Direct input embedding & MAMBA & Huber loss \\
Zezario et al. (2025)~\cite{zezario2025featureimportancedomainsimproving} & Spectral + Waveform + Whisper & MSBG simulation & (Early Attention) CNN + BLSTM + Attention & Utterance-level MSE + Frame-level MSE + Cross Entropy \\
Zhou et al. (2025)~\cite{zhou2025audiogramleveragingexistingscores} & Whisper + Score & None & Transformer & Huber loss \\
\hline
\end{tabular}
}
}
\label{tab:intelligibility_model_comparison}
\end{table*}

\subsection{Listener Audiograms and Acoustic Features}
One of the most important aspects in deploying non-intrusive speech intelligibility prediction for HA is the availability and integration of audiogram information. Audiograms provide essential insights into an individual's hearing thresholds across different frequencies, as shown in Table~\ref{tab:audiogram}, enabling models to personalize intelligibility predictions based on the user's unique hearing profile, as further illustrated in Fig.~\ref{fig:concat}. This information helps ensure that the system accounts for frequency-specific hearing deficits, which is essential for achieving more accurate speech intelligibility estimation.

The use of listener audiograms can generally be categorized into two distinct strategies. The first is direct input embedding, in which the listener’s audiogram is concatenated with the corresponding acoustic features as model input \cite{chiang2021hasa, chiang2023multiobjective, mawalim_isca, cuervo2024speech, zhou2025audiogramleveragingexistingscores}. The second strategy involves incorporating the audiogram into hearing loss simulation models, such as the Moore, Stone, Baer, and Glasberg (MSBG) model \cite{Moore1997, baer1993spectral, baer1994interfering, moore1993loudness} or other ear models \cite{kates_ear}. Several models adopt this simulation-based approach to simulate the perceptual effects of hearing loss and generate the corresponding audio representations \cite{zezario2022mbi, tu22_interspeech, close22_interspeech, 10289742, 9980000, MAWALIM2023109663, 10870355, zezario2025featureimportancedomainsimproving}.

Furthermore, in addition to audiogram information, acoustic features play an important role in accurate intelligibility prediction, given that most datasets \cite{barker2022is, barker2024} consist of challenge scenarios that include unseen speakers, environments, and listener hearing profiles. While handcrafted features such as Mel-Frequency Cepstral Coefficients (MFCC), spectral features, and extended Geneva minimalistic acoustic
parameter set (eGeMAPS) have been shown to be effective features \cite{robach22_interspeech, chiang2021hasa, MAWALIM2023109663, 9980000}, models that incorporate large speech pre-trained models have shown higher prediction performance. These models could either use automatic speech recognition, self-supervised learning (SSL) representations (e.g., HuBERT \cite{hubert}, WavLM \cite{chen2021wavlm}), or weakly supervised models such as Whisper \cite{Whisper}. These models are capable of extracting high-level acoustic and linguistic information that generalizes well across conditions, due to the advantage of being trained on large-scale datasets, thereby improving robustness and intelligibility prediction accuracy. In some scenarios, WavLM tends to achieve higher prediction than HuBERT \cite{zezario2022mbi}, while in more general conditions, Whisper tends to provide richer acoustic features and achieve higher prediction performance \cite{zezario24_interspeech, cuervo2024speech, mogridge2024nonintrusive, zezario2025featureimportancedomainsimproving}. As a result, the combination of personalized hearing profiles (via audiograms) and suitable acoustic features (via SSL or Whisper) has emerged as a key trend in the development of higher-performing and generalizable models.

\subsection{Model Architecture}
Along with richer acoustic features, the choice of a suitable model architecture has also been shown to be a crucial aspect in the development of non-intrusive intelligibility prediction systems. Given the sequential and temporal nature of audio data, most existing methods \cite{chiang2021hasa, chiang2023multiobjective, mogridge2024nonintrusive, zezario2022mbi, zezario24_interspeech, close2023nonintrusiveintelligibilitypredictor, zezario2025featureimportancedomainsimproving} adopt a bidirectional long short-term memory (BLSTM) framework as the backbone of the model. The BLSTM architecture is particularly well-suited for modeling long-range dependencies in audio signals, enabling the system to capture both past and future context, which is essential for accurately predicting speech intelligibility.

To further enhance the modeling capacity of BLSTM-based frameworks, several works have proposed the integration of attention mechanisms \cite{chiang2021hasa, chiang2023multiobjective, close2023nonintrusiveintelligibilitypredictor, mogridge2024nonintrusive, 10870355}. These mechanisms allow the model to focus on more important regions of the input sequence, thereby improving its ability to respond to important acoustic or linguistic cues. Another line of research introduces CNN into the BLSTM architecture \cite{zezario2022mbi, zezario24_interspeech, zezario2025featureimportancedomainsimproving}, as front-end feature extractors. The use of CNN helps capture local spectro-temporal patterns that strengthen the sequence modeling performed by BLSTM.

Beyond the BLSTM-based approaches, several alternative architectures have been explored to improve performance or reduce model complexity. For instance, time-delay neural networks (TDNN) \cite{robach22_interspeech} have been investigated for their ability to model temporal context with fewer parameters and lower computational cost. CNN-based architectures \cite{close22_interspeech, MAWALIM2023109663, 9980000} have also been proposed, leveraging deep hierarchical feature representations. More recently, transformer-based models \cite{tu22_interspeech, cuervo2024speech, zhou2025audiogramleveragingexistingscores} have gained attention due to their ability to capture global dependencies and their success in other speech-related tasks. Furthermore, the emerging MAMBA architecture \cite{yamamoto2025nonintrusivebinauralspeechintelligibility}, designed to handle long sequences, presents an alternative module for non-intrusive speech intelligibility prediction in HA.

\subsection{Objective Functions}
To optimally train the model, it is crucial to define an appropriate objective function that effectively guides the adjustment of model parameters. A commonly adopted approach is direct estimation, which minimizes the mean square error (MSE) between the model’s predicted scores and the corresponding human-labeled intelligibility scores \cite{close22_interspeech, MAWALIM2023109663}, as defined in the following equation:
\begin{equation}
\mathcal{L}_{\text{utterance}} = \frac{1}{N} \sum_{i=1}^{N} \left(y_i-\hat{y}_i\right)^2
\end{equation}
where \( y_i \) and \( \hat{y}_i \) denote the reference and predicted intelligibility scores for the \( i \)-th utterance.

In addition to utterance-level loss, several studies have explored the benefits of incorporating frame-level objectives as auxiliary losses. This approach aims to capture more fine-grained temporal patterns during training, which may improve generalization and stabilize learning \cite{chiang2021hasa, zezario2022mbi, zezario24_interspeech, chiang2023multiobjective, zezario2025featureimportancedomainsimproving}.
\begin{equation}
\mathcal{L}_{\text{frame}} = \frac{1}{N} \sum_{i=1}^{N} \frac{1}{T_i} \sum_{t=1}^{T_i} \left(y_{i,t}-\hat{y}_{i,t} \right)^2
\end{equation}
where \( T_i \) is the number of frames in the \( i \)-th utterance. In general, prior works \cite{chiang2021hasa, zezario2022mbi, zezario24_interspeech, chiang2023multiobjective, zezario2025featureimportancedomainsimproving} combine \(\mathcal{L}_{\text{utterance}}\) and \(\mathcal{L}_{\text{frame}}\) to form the overall training loss.

Some studies \cite{zezario24_interspeech, zezario2025featureimportancedomainsimproving} further adopt a multi-task learning strategy, combining losses from different metrics such as HASPI, along with a cross-entropy loss to classify different HA systems, as follows:
\begin{equation}
\mathcal{L}_{\text{multi-task}} = \alpha \cdot \mathcal{L}_{\text{intell}} + (1 - \alpha) \cdot \mathcal{L}_{\text{HASPI}} + \mathcal{L}_{\text{CE}}
\end{equation}
where \( \alpha \in [0, 1] \) controls the task weighting. Each main loss (\(\mathcal{L}_{\text{intell}}\) and \(\mathcal{L}_{\text{HASPI}}\)) is defined as a weighted combination of utterance-level and frame-level losses:
\begin{align}
\mathcal{L}_{\text{intell}} &= \beta \cdot \mathcal{L}_{\text{intell,utt}} + (1 - \beta) \cdot \mathcal{L}_{\text{intell,frame}} \\
\mathcal{L}_{\text{HASPI}} &= \beta \cdot \mathcal{L}_{\text{HASPI,utt}} + (1 - \beta) \cdot \mathcal{L}_{\text{HASPI,frame}}
\end{align}

The individual loss terms are defined as:
\begin{align}
\mathcal{L}_{\text{intell,utt}} &= \frac{1}{N} \sum_{i=1}^{N} \left( y^{\text{intell}}_{i}- \hat{y}^{\text{intell}}_{i} \right)^2 \\
\mathcal{L}_{\text{intell,frame}} &= \frac{1}{N} \sum_{i=1}^{N} \frac{1}{T_i} \sum_{t=1}^{T_i} \left(y^{\text{intell}}_{i,t}- \hat{y}^{\text{intell}}_{i,t}  \right)^2 \\
\mathcal{L}_{\text{HASPI,utt}} &= \frac{1}{N} \sum_{i=1}^{N} \left( y^{\text{HASPI}}_{i}-\hat{y}^{\text{HASPI}}_{i} \right)^2 \\
\mathcal{L}_{\text{HASPI,frame}} &= \frac{1}{N} \sum_{i=1}^{N} \frac{1}{T_i} \sum_{t=1}^{T_i} \left(y^{\text{HASPI}}_{i,t}- \hat{y}^{\text{HASPI}}_{i,t}  \right)^2
\end{align}

Moreover, alternative loss formulations such as the Huber\cite{huber} loss have also been investigated for their robustness to outliers and their balanced treatment between L1 and L2 behaviors \cite{cuervo2024speech, yamamoto2025nonintrusivebinauralspeechintelligibility, zhou2025audiogramleveragingexistingscores}:
\begin{equation}
\mathcal{L}_{\text{Huber}} = \frac{1}{N} \sum_{i=1}^{N} 
\begin{cases}
\frac{1}{2} \left(y_i-\hat{y}_i  \right)^2 & \text{if } \left| \hat{y}_i - y_i \right| \leq \delta \\
\delta \left( \left| y_i -\hat{y}_i \right| - \frac{1}{2} \delta \right) & \text{otherwise}
\end{cases}
\end{equation}
where \( \delta \) is a threshold parameter.

In a slightly different scenario, where ASR-based systems are used to extract confidence measures, the Connectionist Temporal Classification (CTC) loss is often selected as the training objective.

\subsection{Dataset}
Due to the data-driven nature of non-intrusive speech intelligibility prediction methods, the availability of suitable datasets is an important factor in system development. The Clarity Prediction Challenge (CPC) series—CPC1 \cite{barker2022is} and CPC2 \cite{barker2024}—plays an important role in advancing this field, with most existing deep learning-based approaches for HA assessment relying on these datasets for training and evaluation.

CPC1 (2022) \cite{barker2022is} provides data from ten HA systems used in the 2021 Clarity Enhancement Challenge \cite{clarity}. Listening tests involve 25 HA users, each tasked with repeating what they hear from presented speech samples. Scores range from 0 to 100, with higher scores indicating better intelligibility. Bilateral pure-tone audiograms are estimated for each listener using hearing thresholds at [250, 500, 1000, 2000, 3000, 4000, 6000, 8000] Hz. The dataset is organized into two tracks: Track 1 with 4,863 training utterances and Track 2 with 3,580 training utterances, alongside corresponding test sets of 2,421 and 632 utterances, respectively, with no overlap between training and test material.

Building upon CPC1, CPC2 (2023) \cite{barker2024} expands the dataset to include recordings from six talkers processed by ten HA systems originating from the 2022 Clarity Enhancement Challenge \cite{10094918}. Intelligibility ratings from 25 listeners are collected for three tracks: Track 1 (2,779 utterances), Track 2 (2,796 utterances), and Track 3 (2,772 utterances). The test sets—containing 305, 294, and 298 utterances for Tracks 1–3, respectively—feature unseen listeners and HA systems. Evaluation relies on root mean square error (RMSE), linear correlation coefficient (LCC), and Spearman’s rank correlation coefficient (SRCC) \cite{srcc}, where lower RMSE and higher LCC/SRCC indicate better performance.

\section{Key Challenges}

Based on recent advancements, several key challenges remain in the development of non-intrusive intelligibility prediction models for HA users. One of the most important challenges is the effective integration of a listener’s hearing loss audiogram. While many models incorporate audiogram data by directly concatenating it with acoustic features, this naïve embedding approach may not fully capture the complex perceptual information of hearing impairment. On the other hand, integrated simulation-based hearing loss models such as MSBG and EarModel may introduce additional complexity, can often be non-differentiable, and may limit the efficiency of end-to-end learning. Furthermore, in scenarios where the systems are tested on new listener data, the systems may struggle to generalize well if they were previously trained on specific listener audiograms. As a result, recent developments mainly focus on improving generalization under more adverse conditions. With the progression from  CPC 1 \cite{barker2022is} and CPC2 \cite{barker2024} to the current CPC3 dataset, the emphasis has shifted toward using general severity levels as substitutes for listener-specific audiograms. This highlights a clear trend toward developing more generalizable systems capable of operating effectively under diverse and challenging conditions.

Another fundamental problem is ensuring robustness and generalizability across diverse acoustic environments. Many existing systems perform well on seen datasets but degrade significantly in real-world scenarios involving unseen speakers, noise types, room acoustics, and hearing-aid settings. Additionally, the selection of acoustic features also plays a pivotal role in model performance. Earlier approaches primarily relied on handcrafted features such as Mel-Frequency Cepstral Coefficients (MFCC) or other spectral features, which are computationally efficient but limited in modeling long-range temporal patterns and higher-order contextual dependencies. More recent works have incorporated self-supervised learning (SSL) models such as HuBERT, WavLM, and Whisper, which provide rich and hierarchical representations learned from large-scale speech corpora. Although these features improve prediction accuracy, they also make the model more computationally intensive. Therefore, developing appropriate strategies to reduce computational cost while maintaining generalization remains an important step for better system deployment.

Lastly, there is a need to design objective functions that better align with human perception. While most models optimize for mean squared error (MSE) between predicted and ground-truth intelligibility scores, this metric may not adequately capture perceptual information. Recent efforts have explored auxiliary frame-level losses or robust alternatives like the Huber loss, but perceptually inspired or listener-specific loss functions remain limited.

In summary, future progress in non-intrusive intelligibility prediction will depend on improving hearing loss modeling in ways that are both perceptually meaningful and computationally feasible, enhancing robustness to unseen conditions, leveraging rich yet efficient acoustic representations, and aligning training objectives more closely with human-perceived intelligibility.

\section{Conclusion and Discussions}
In this paper, we reviewed recent developments in deep learning-based non-intrusive intelligibility prediction for HA. A variety of input features have been explored, ranging from traditional acoustic features like MFCC to self-supervised embeddings such as Whisper and WavLM. At the same time, hearing loss modeling has been further explored from simple audiogram concatenation to more perceptually accurate simulations using models like MSBG. Architecturally, researchers have transitioned from CNNs and BLSTMs to more advanced structures, including attention mechanisms, transformers, and memory-based designs. Despite these improvements, most models are still trained with loss functions such as MSE or Huber loss, which may not fully capture the perceptual nature of intelligibility as rated by hearing-impaired listeners.

Looking forward, future development of non-intrusive intelligibility prediction models for HA may benefit from incorporating hearing loss modeling approaches that balance perceptual relevance with computational efficiency. This may involve the design of learnable and differentiable auditory models, or hybrid strategies that combine severity-based representations with listener-specific adaptation to improve generalization under mismatched audiogram conditions. Enhancing robustness to unseen acoustic scenarios will also be important, potentially achieved through domain generalization techniques, data augmentation, and model architectures that can adapt to varying noise and reverberation conditions. Furthermore, balancing prediction accuracy with computational efficiency emphasizes the value of lightweight feature representations that preserve the contextual richness of SSL features while supporting real-time deployment. Finally, exploring perceptually guided and listener-aware loss functions, possibly within multi-task learning frameworks, may further align model optimization with human-perceived intelligibility and enhance practical applicability in HA scenarios.

\printbibliography[title=References]
\end{document}